\documentclass[onecolumn,prd,showpacs,showkeys,preprintnumbers,amsmath,amssymb]{revtex4}

\newcommand{\beq}{\begin{equation}}
\newcommand{\eeq}{\end{equation}}

\newcommand{\bfxi}{\mbox{\boldmath $\xi$}}

\newcommand{\bfx}{{\bf x}}
\def\bfv{{\bf v}}

\def\gs{\mathrel{\lower0.6ex\hbox{$\buildrel {\textstyle >}\over{\scriptstyle \sim}$}}}
\def\ls{\mathrel{\lower0.6ex\hbox{$\buildrel {\textstyle <}\over{\scriptstyle \sim}$}}}

\begin{document}

\title{Two-body problem with the cosmological constant and observational constraints}

\author{Philippe Jetzer}
\email{jetzer@physik.unizh.ch}
\author{Mauro Sereno}
\email{sereno@physik.unizh.ch}
\affiliation{Institut f\"{u}r Theoretische Physik, Universit\"{a}t Z\"{u}rich,
Winterthurerstrasse 190, CH-8057 Z\"{u}rich , Switzerland}

\date{November 21, 2005}

\begin{abstract}
We discuss the influence of the cosmological constant on the
gravitational equations of motion of bodies with arbitrary masses and eventually
solve the two-body problem. Observational constraints are derived from
measurements of the periastron advance in stellar systems, in
particular binary pulsars and the solar system. Up to now, Earth and Mars data
give the best constraint, $\Lambda~\ls~10^{-36}\mathrm{km}^{-2}$;
bounds from binary pulsars are potentially competitive with limits
from interplanetary measurements. If properly accounting for the
gravito-magnetic effect, this upper limit on $\Lambda$ could greatly
improve in the near future thanks to new data from planned or already
operating space-missions.
\end{abstract}

\maketitle

\section{Introduction}

Aged nearly one century, Einstein's cosmological constant $\Lambda$
still keeps unchanged its cool role to solve problems. $\Lambda$,
despite being just one number, was able to respond to very different
needs of the scientific community, from theoretical prejudices about
universe being static (which provided the original motivation for
introducing $\Lambda$ in 1917) to observational hints that the
universe is dominated by unclustered energy density exerting negative
pressure, as required by data of exquisite quality which became
available in the last couple of decades. Although it is apparently
plagued by some theoretical problems about its size and the
coincidence that just in the current phase of the universe the energy
contribution from $\Lambda$ is of the same order of that from
non-relativistic matter, the cosmological constant still provides the
most economical and simplest explanation for all the cosmological
observations \cite{pad05}. The interpretation of the cosmological
constant is a very fascinating and traditional topic.
$\Lambda$ might be connected to the vacuum density, as suggested by
various authors (see \cite{pe+ra03} for an historical account), and could offer the greatest contribution from cosmology to fundamental physics.

The big interest in the cosmological constant has recursively raised
attempts in putting observational bounds on its absolute value from
completely different phenomena. $\Lambda$, supposed to
be $\sim 10^{-46}\mathrm{km}^{-2}$ from observational cosmology analyses, is obviously of relevance on cosmological scales but it could play some role also in local
problems. Up to now, no convincing methods for constraining $\Lambda$
in a laboratory have been proposed \cite{je+st05}, but interesting
results have been obtained considering planetary motions in the solar
system \cite{isl83,wri98,ker+al03}. The effects of $\Lambda$ become
stronger for diluted mass conglomeration but they get enhanced also
through various mechanisms \cite{now+al02,ba+no05}. As an example,
conditions for the virial equilibrium can be affected by $\Lambda$ for
highly flattened objects \cite{now+al02}. On the scale of the Local Volume, a cosmological constant could have observable consequences by producing lower velocity dispersion around the Hubble flow \cite{tee+al05}.

Up to now, local physical consequences of the existence of a
cosmological constant were investigated studying the motion of test
bodies in the gravitational field of a very large mass. This one-body
problem can be properly considered in the framework of the spherically
symmetric Schwarzschild vacuum solution with a cosmological constant,
also known as Schwarzschild-de Sitter or Kottler space-time. The rotation of the central source can also be
accounted for using the so-called Kerr-de Sitter space-time
\cite{ker+al03}. Here, we carry out an analysis of the gravitational
$N$-body problem with arbitrary masses in the weak field limit with a
cosmological constant. This study is motivated by the more and more
central role of binary pulsars, from the discovery of the pulsar PSR
B1913+16 in 1974 \cite{hu+ta75}, in testing gravitational and relativistic
effects. The gravitational two-body equations of motion for arbitrary
masses were first derived in absence of spin by Einstein, Infeld and
Hoffmann (EIH) \cite{ein+al38}. The problem was later addressed in
more general cases, subsequently accounting for spins and quadrupole
moments \cite[and reference therein]{ba+oc75}. Here, we take the
further step to consider a cosmological constant.

The paper is as follows. In section~\ref{sec:field}, we discuss the
gravitational weak field limit in presence of a
cosmological constant and introduce the relevant approximations.
Section~\ref{sec:EIHeq} presents the generalization of the EIH equations of motion,
whereas section~\ref{sec:twobo} is devoted to the
study of the two-body problem. In section~\ref{sec:obser}, we review
how measurements of precession of pericentre in stellar system can
constrain $\Lambda$. In particular, we consider binary pulsars and the
solar system. Section~\ref{sec:concl} contains some final
considerations.

\section{Field equations with a cosmological constant in post-Newtonian approximation}
\label{sec:field}

Einstein's equations with the cosmological constant are
\begin{equation}
R_{\mu \nu} - \Lambda g_{\mu \nu} = \frac{8 \pi G}{c^4} S_{\mu \nu} ,
\end{equation}
where $G$ is the gravitational constant, $c$ the vacuum speed of light
and
\begin{equation}
S_{\mu \nu} \equiv  T_{\mu \nu} - \frac{1}{2} g_{\mu \nu}
T^{\lambda}_{\lambda},
\end{equation}
with $ T_{\mu \nu}$ being the energy-momentum tensor. The weak field expansion can start by introducing a nearly Lorentzian
system for weak, quasi-stationary fields, in which
\begin{equation}
g_{\mu \nu} = \eta_{\mu \nu} + h_{\mu \nu}, \ \ \ |h_{\mu \nu}| \ll 1.
\end{equation}
Actually, the Minkowski metric $\eta_{\mu \nu}$ is not a vacuum
solution of the field equations with a cosmological constant, but for
$ | \Lambda |  \ll 1$ an approximate solution in a finite region can
still be found by an expansion around $\eta_{\mu \nu}$. In the
post-Newtonian (pN) approximation, metric components can be expanded
in powers of
\begin{equation}
\varepsilon \sim \left( \frac{G M}{c^2 R} \right)^{1/2} \sim \frac{v}{c} \sim
\frac{p}{\rho},
\end{equation}
where $M$, $R$, $v$, $p$ and $\rho$ represent typical values for the mass,
length, velocity, pressure and energy density of the system, respectively. In what follows, $^{(n)}g_{\mu \nu}$ and $^{(n)}T_{\mu \nu}$ will
denote terms of order $\varepsilon^n$ and $\varepsilon^n (M/R)$,
respectively. To perform a proper treatment in presence of a $\Lambda$-term we have to consider the suitable approximation order for $\Lambda$. We assume that the size of the contributions due to the cosmological constant is at most comparable to the post-Newtonian terms, i.e, ${\cal{O}}(\Lambda g_{00}) \gs {\cal{O}} (G\ ^{(2)} T^{00}/c^2)$. This condition can be rewritten as 
\begin{equation}
\label{fie1a}
\Lambda  \ls  \frac{R_\mathrm{g}^2}{R^4} ,
\end{equation}
where $R_\mathrm{g} \equiv G M/c^2$ is the gravitational radius. Eq.~(\ref{fie1a}) is easily satisfied by gravitational bound systems with $M \sim M_\odot$ and $ R\sim 1-10^2~\mathrm{AU}$ if $\Lambda \ls 10^{-33}~\mathrm{km}$, a value well above the estimated one from cosmological constraints and also greater than the limits we will derive in section \ref{sec:obser} considering stellar systems. Hereafter, we will put $c=1$. With such an approximation order, we can use classical results within the standard pN gauge. Following
\citet{str04}, the approximate field equations read
\begin{eqnarray}
\Delta^{(2)} g_{00} & = & -8 \pi G \ ^{(0)} T^{00},
\\
\Delta^{(4)} g_{00} & = & \ ^{(2)}g_{ij}\ ^{(2)}g_{00,ij}+
^{(2)}g_{ij,j}\ ^{(2)} g_{00,i}
- \frac{1}{2}\ ^{(2)} g_{00,i}\ ^{(2)} g_{00,i}
- \frac{1}{2}\ ^{(2)} g_{00,i}\ ^{(2)} g_{jj,i}
\\
& -&  8 \pi G \left(\ ^{(2)} T^{00}-2^{\ (2)}g_{00}\ ^{(0)}T^{00}
+^{\ (2)}T^{ii} \right) +2 \Lambda , \nonumber
\\
\Delta^{(3)} g_{oi} & = & - \frac{1}{2}\ ^{(2)}g_{jj,0i}+
^{\ (2)}g_{ij,0j} + 16 \pi G ^{\ (1)} T^{i0} ,
\\
\Delta^{(2)} g_{ij} & = & - 8 \pi G \delta_{ij} \ ^{(0)} T^{00} .
\end{eqnarray}
The components of the metric can be expressed in terms of potentials.
Let $\phi_\mathrm{N}$ be the Newtonian potential,
\begin{equation}
\phi_\mathrm{N} = -G \int \frac{^{(0)} T^{00} (t,\bfx^{'})}{|\bfx-\bfx^{'}|}d^3 x^{i} .
\end{equation}
According to our approximation order, the cosmological constant
appears only in the equation for $^{(4)} g_{00}$. This can be
re-arranged to give
\begin{equation}
\Delta^{(4)} ( g_{00} +2 \phi_\mathrm{N}^2 )=  -8 \pi G \left(\ ^{(2)}
T^{00} + ^{(2)} T^{ii} \right) +2 \Lambda
\end{equation}
Together with the classical pN potential $\psi$,
\begin{equation}
\psi = -G \int \frac{d^3 x^{'}}{|\bfx-\bfx^{'}|} \left( \ ^{(2)}T^{00}+
^{(2)} T^{ii} \right) ,
\end{equation}
we introduce $\phi_\Lambda$, solution of
\begin{equation}
\Delta \phi_\Lambda = -\Lambda .
\end{equation}
In presence of a cosmological constant, there is an upper limit on the
maximum distance within which the Newtonian limit holds and boundary
conditions must then be chosen at a finite range \citep{now01}. When
these boundary conditions are chosen on a sphere whose origin
coincides with the origin of the coordinate system, $\phi_\Lambda$ can
be expressed as
\begin{equation}
\phi_\Lambda = -\frac{1}{6} \Lambda |\bfx|^2 ,
\end{equation}
where we have neglected correction terms which appear because of
boundary conditions. Due to a positive cosmological constant, the origin of the coordinate system has a distinguished dynamical role with a radial force directed away from it
\cite{adl+al65}. Since the choice of the origin is arbitrary, any point in the
space will experience repulsion
from any other point. Finally, introducing the standard pN potentials,
\begin{eqnarray}
\xi_i & = & - 4 G \int \frac{d^3 x^{'}}{|\bfx-\bfx^{'}|}\ ^{(1)} T^{i0}
(t,\bfx^{'}), \\
\chi & = & - \frac{G}{2} \int   |\bfx-\bfx^{'}|  ^{(0)} T^{00} d^3 x^{'}
(t,\bfx^{'}),
\end{eqnarray}
the metric components read
\begin{eqnarray}
^{(2)}g_{00} & = & -2 \phi_\mathrm{N} , \label{metr1} \\
^{(4)}g_{00} & = & -2 ( \phi_\mathrm{N}^2 +\psi + \phi_\Lambda ), \label{metr2}  \\
^{(2)}g_{ij} & = & -2 \delta_{ij} \phi_\mathrm{N} , \label{metr3} \\
^{(3)}g_{0i} & = & \xi_i +\chi_{,i0} . \label{metr4}
\end{eqnarray}
For a point-like mass at the centre of the coordinate system, the
above expressions reduce to the weak field limit at large radii of the
Kottler space-time.

\subsection{Equations of motion for a test particle}

The motion of a particle in an external gravitational field can be
described by the Lagrangian
\begin{equation}
{\cal L}= 1- \sqrt{ - g_{\mu \nu} \left( \frac{d x^{\mu}}{dt} \right)
\left( \frac{d x^{\nu}}{dt} \right)}.
\end{equation}
Using the metric components in equations~(\ref{metr1}-\ref{metr4}), we get
\begin{equation}
{\cal {L}} \simeq \frac{1}{2} v^2+ \frac{1}{8} v^4 -\phi_\mathrm{N} -\frac{1}{2}
\phi_\mathrm{N}^2 -\psi -\phi_\Lambda -\frac{3}{2}\phi_\mathrm{N} v^2 +v^i \left( \xi_i +
\frac{\partial \chi}{\partial t \partial x^i} \right) .
\end{equation}
The corresponding Euler-Lagrange equations of motion in a
3-dimensional notation read,
\begin{equation}
\frac{d {\bf v}}{dt} \simeq -\nabla \left( \phi_\mathrm{N} +2\phi_\mathrm{N}^2+\psi \right) -
\frac{\partial \bfxi}{\partial t} - \frac{\partial^2 }{\partial t^2} \nabla \chi +
\bfv {\times} (\nabla {\times} \bfxi) + 3 \bfv \frac{\partial \phi_\mathrm{N}}{\partial t}
+4 \bfv (\bfv {\cdot} \nabla)\phi_\mathrm{N}-\bfv^ 2\nabla \phi_\mathrm{N} +\frac{\Lambda}{3}
\bfx .
\end{equation}
The above expression reduces to equation (20) in \cite{ker+al03} when
neglecting pN corrections.

\section{The Einstein-Infeld-Hoffmann equations}
\label{sec:EIHeq}

Since the contribution from the cosmological constant is of higher-order,
it does not couple with other corrections. The Lagrangian of an $N$-body
system of point-like particles can be written as
\begin{equation}
{\cal L} \simeq {\cal L}_{(\Lambda =0)} + \delta{\cal L}_\Lambda,
\end{equation}
where ${\cal L}_{\Lambda =0}$ is the total Lagrangian in absence of
$\Lambda$. The Lagrangian ${\cal L}_a $ of particle $a$ in the field of
other particles is
\begin{equation}
{\cal L}_a \simeq {\cal L}_{a(\Lambda =0)} +\frac{\Lambda}{6} \bfx_a^2,
\end{equation}
where ${\cal L}_{a(\Lambda =0)}$ is given in equation (5.94) in
\cite{str04} The total Lagrangian reads
\begin{equation}
{\cal L} \simeq {\cal L}_{(\Lambda =0)} + \sum_a \frac{\Lambda}{6} m_a
\bfx_a^2,
\end{equation}
with ${\cal L}_{(\Lambda =0)}$ given in \cite[equation~(5.95)]{str04}.
The corresponding Euler-Lagrange equations are the Einstein-Infeld-Hoffmann equations corrected for a $\Lambda$ term,
\begin{equation}
\dot{\bfv}_a = - G \sum_{b \neq a} \left( \frac{\bfx_{ab}}{r_{ab}} \right)+
\delta \mathbf{F}_{\mathrm{pN} (\Lambda=0)} + \frac{\Lambda}{3} \bfx_a
\end{equation}
where $\mathbf{F}_{\mathrm{pN} (\Lambda=0)}$ is the post-Newtonian perturbing
function \cite[equation~(5.96)]{str04}.

\section{The two-body problem}
\label{sec:twobo}

The total Lagrangian for two particles can be written as
\begin{equation}
{\cal L} \simeq \frac{1}{2}m_\mathrm{a} v_\mathrm{a}^2 +G \frac{m_\mathrm{a} m_\mathrm{b}}{x}+ \frac{1}{2}m_\mathrm{b}
v_\mathrm{b}^2 + \delta {\cal L}_{\mathrm{pN} (\Lambda=0)} + \delta {\cal L}_{\Lambda}
\end{equation}
where $\bfx  \equiv \bfx_\mathrm{a} - \bfx_\mathrm{b} $ is the separation vector and $\delta {\cal L}_{\mathrm{pN} (\Lambda=0)}$ and  $\delta {\cal L}_{\Lambda}$ are the pN and $\Lambda$-contributions, respectively. It is \cite{str04}
\begin{equation}
\delta {\cal
L}_{\mathrm{pN} (\Lambda=0)} = \frac{1}{8} (m_\mathrm{a} v_\mathrm{a}^4 + m_\mathrm{b} v_\mathrm{b}^4) + G
\frac{m_\mathrm{a} m_\mathrm{b}}{2 r} \left[ 3 (v_\mathrm{a}^2+v_\mathrm{b}^2) - 7 \bfv_\mathrm{a} \dot \bfv_\mathrm{b} -
(\bfv_\mathrm{a} \dot {\bf n})(\bfv_\mathrm{b} \dot {\bf n}) \right]
-\frac{G^2}{2}\frac{m_\mathrm{a} m_\mathrm{b} (m_\mathrm{a}+m_\mathrm{b})}{x^2}
\end{equation}
with ${\bf n} \equiv {\bf x}/x$ and
\begin{equation}
\delta {\cal L}_{\Lambda} = \frac{\Lambda}{6} (m_\mathrm{a} x_\mathrm{a}^2 + m_\mathrm{b} x_\mathrm{b}^2) .
\end{equation}
Due to cosmological constant, the energy of the system is modified by
a contribution $-\delta {\cal L}_{\Lambda}$. The pN and $\Lambda$
corrections are additive and can be treated separately. We are
interested in examining the effect of a non vanishing $\Lambda$ term.
Let us consider the centre of mass and relative motions. Introducing
${\bf X} \equiv  \left( m_\mathrm{a} \bfx_\mathrm{a} +m_\mathrm{b}
\bfx_\mathrm{b} \right)/M $, with $M \equiv
m_\mathrm{a}+m_\mathrm{b}$, the Lagrangian can be re-written as
\begin{equation}
{\cal L} \simeq \frac{1}{2} M V^2 + \frac{\Lambda}{6} M X^2 +\frac{1}{2} \mu
v^2 + \frac{\Lambda}{6} \mu x^2  + G \frac{M \mu}{x},
\end{equation}
with $\mu \equiv m_\mathrm{a} m_\mathrm{b}/M$. Due to cosmological constant, the centre of mass of the
system is subject to an effective repulsive force given by $\Lambda {\bf X}/3$ per unit mass.

The equations for the relative motion are those of a test particle in
a Schwarzschild-de~Sitter space-time with a source mass equal to the
total mass of the two-body system. Since the perturbation due to
$\Lambda$ is radial, the orbital angular momentum is conserved and the
orbit is planar. The main effect of $\Lambda$ on the orbital motion is
a precession of the pericentre \cite[and references
therein]{kr+wh03,ker+al03}. Following the analysis of the Rung-Lenz
vector in \cite{ker+al03} and restoring the $c$ factors, we get for the contribution to the precession angular velocity due to $\Lambda$,
\begin{equation}
\dot{\omega}_\Lambda = \frac{\Lambda c^2 P_\mathrm{b}}{4 \pi}
\sqrt{1-e^2},
\end{equation}
where $e$ is the eccentricity and $P_\mathrm{b}$ the Keplerian period of
the unperturbed orbit. This contribution should be considered together with the post-Newtonian
periastron advance, $\dot{\omega}_{\mathrm{pN}} = 3 (2 \pi/P_\mathrm{b})^{5/3} (G
M/c^3)^{2/3} (1-e^2)^{-1} $. The ratio between these two contributions
can be written as,
\begin{equation}
\frac{ \dot{\omega}_\Lambda } {\dot{\omega}_{\mathrm{pN}}} =
\frac{\bar{R}}{R_\mathrm{g}} \frac{\rho_\Lambda }{ \rho } 
= \frac{1}{6}\frac{\bar{R}^4}{R_\mathrm{g}^2} \Lambda  ,
\end{equation}
where $\bar{R} = a (1-e^2)^{3/8}$ is a typical orbital radius with $a$ the semi-major radius of the unperturbed orbit, $\rho \equiv M/(4 \pi \bar{R}^3/3)$ is a typical density of the system and $\rho_\Lambda
\equiv c^2 \Lambda /8 \pi G $ is the energy density associated to
the cosmological constant. The effect of $\Lambda$ can be significant for
very wide systems with a very small mass.

\section{Observational constraints}
\label{sec:obser}

In this section, we derive observational limits on $\Lambda$ from
orbital precession shifts in stellar systems and in the solar system.

\subsection{Interplanetary measures}
\label{sec:obserA}

Precessions of the perihelia of the solar system planets have provided
the most sensitive local tests for a cosmological constant so far
\cite{isl83,wri98,ker+al03}. Estimates of the anomalous perihelion
advance were recently determined for Mercury, Earth and Mars \cite{pit05a,pit05b}. Such ephemerides were constructed integrating
the equation of motion for all planets, the Sun, the Moon and largest
asteroids and including rotations of the Earth and of the Moon,
perturbations from the solar quadrupole mass moment and asteroid ring
in the ecliptic plane. Extra-corrections to the known general relativistic predictions can
be interpreted in terms of a cosmological constant effect. We considered the 1-$\sigma$ upper bounds. Results are listed in Table~\ref{tab:plan}. Best constraints come from Earth and
Mars observations, with $\Lambda \ls 10^{-36}\mathrm{km}^{-2}$. Major sources of systematic errors come from uncertainties about solar oblateness and from the
gravito-magnetic contribution to secular advance of perihelion but
their effect could be in principle accounted for \citep{ior05}. In
particular, the general relativistic Lense-Thirring secular precession
of perihelia is compatible with the determined extra-precessions
\citep{ior05}. The accuracy in determining the planetary orbital
motions will further improve with data from space-missions like
BepiColombo, Messenger and Venus express. By considering a
post-Newtonian dynamics inclusive of gravito-magnetic terms, the
resulting residual extra-precessions should be reduced by several
orders of magnitude, greatly improving the upper bound on $\Lambda$.

The orbital motion of laser-ranged satellites around the Earth has been also considered to confirm general relativistic predictions. Observations of the rates of change of the nodal longitude of the LAGEOS satellites allowed to probe the Lense-Thirring effect with an accuracy of $\sim10\%$, i.e. about half a milliarcsecond per year \cite{ci+pa04}. Other proposed missions, such as the LARES/WEBER-SAT satellite \cite{ciu04}, should further increase this experimental precision. In general, since effects of $\Lambda$ become significant only for very dilute systems, even very accurate measurements of orbital elements of Earth's satellites can not help in constraining the cosmological constant. For a satellite with a typical orbital semi-major axis of about 12,000~km, in order to get a bound on $\Lambda$ as accurate as those inferred from Earth and Mars perihelion shifts (i.e. $\Lambda \ls 10^{-36}~\mathrm{km}^{-2}$), changes in orbital elements should be measured with a today unattainable precision of a few tens of picoseconds of arc per year, about six order of magnitude better than today accuracy.

\begin{table}
\caption{\label{tab:plan} Limits on the cosmological constant
due to extra-precession of the inner planets of the solar system.}
\begin{ruledtabular}
\begin{tabular}{lrrr}
Name  &  $\delta\dot{\omega}\footnotemark[1]$ (arcsec/year)&
 $\dot{\omega}_\Lambda$ ($\deg$/year)& $\Lambda_\mathrm{lim}~(\mathrm{km}^{-2}) $ \\
\hline
Mercury  &  $-0.36(50)\times 10^{-4}$   &  $9.61{\times} 10^{25} \Lambda /(1~\mathrm{km}^{-2})$ &
$4 {\times} 10^{-35}$
\\
Venus    &  $0.53(30)\times 10^{-2}$     &  $2.51{\times} 10^{26}\Lambda /(1~\mathrm{km}^{-2})$ &
$9 {\times} 10^{-33}$
\\
Earth    &  $-0.2(4)\times 10^{-5}$  & $4.08{\times} 10^{26}\Lambda /(1~\mathrm{km}^{-2})$ &   $1{\times}
10^{-36}$
\\
Mars &      $0.1(5)\times 10^{-5}$  & $7.64{\times} 10^{26} \Lambda /(1~\mathrm{km}^{-2})$& $ 2{\times}
10^{-36}$
\\
\end{tabular}
\end{ruledtabular}
\footnotetext[1]{From \cite{pit05a}.}
\end{table}

\subsection{Binary pulsars}

\begin{table}
\caption{\label{tab:1} Binary pulsars with known post-Keplerian parameter
$\dot{\omega}$ and corresponding limits on the cosmological constant.
The identification of the companion is often uncertain. We refer to
the original papers for a complete discussion.}
\begin{ruledtabular}
\begin{tabular}{lllllcc}
PSR Name  &  $P_\mathrm{b}$ (days) &  $e$  &  $\dot{\omega}$ ($\deg$/year)&
 $\dot{\omega}_\Lambda$ ($\deg$/year)& $ \Lambda_\mathrm{lim}~(\mathrm{km}^{-2})$ & ref.\\
\hline
\multicolumn{7}{c}{Double neutron star binaries} \\
\hline
J1518+4904  & 8.634000485     & 0.2494849       &       0.0111(2) &
$9.335 {\times} 10^{24}\Lambda /(1~\mathrm{km}^{-2})$ & $2 {\times} 10^{-29}$ &
\cite{nic+al96}
\\
B1534+12    &       0.2736767       &       0.420737299153  &
1.755805(3) &       $2.772 {\times} 10^{23}\Lambda /(1~\mathrm{km}^{-2})$ & $1 {\times}
10^{-29}$  & \cite{wil05}   \\
B1913+16    &       0.323           &
0.617 & 4.226595(5) &       $2.838{\times} 10^{23} \Lambda /(1~\mathrm{km}^{-2})$       &
$2
{\times} 10^{-29}$ & \cite{wil05}
\\
J1756-2251 & 0.319633898     & 0.180567        & 2.585(2) & $3.510
{\times} 10^{23} \Lambda /(1~\mathrm{km}^{-2})$  & $6 {\times} 10^{-27}$  & \cite{fau+al05}
\\
J1811-1736      &       18.779168       &       0.82802 &       0.009(2)  &
 $1.176{\times} 10^{25} \Lambda /(1~\mathrm{km}^{-2})$      &       $2 {\times} 10^{-28}$ & \cite{lyn+al00}
\\
J1829+2456  & 1.176028 & 0.13914 &       0.28(1) & $1.300 {\times}
10^{24} \Lambda /(1~\mathrm{km}^{-2})$ & $ 8 {\times} 10^{-27}$ & \cite{cha+al04}
\\
B2127+11C   & 0.68141 & 0.335282052     &       4.457(12) &
$7.168{\times} 10^{23}\Lambda /(1~\mathrm{km}^{-2})$ & $2{\times} 10^{-26}$  & \cite{wil05}
\\
B2303+46    & 12.34 & 0.65837 &       0.01019(13)         & $1.037 {\times}
10^{25}  \Lambda /(1~\mathrm{km}^{-2}) $     & $1 {\times} 10^{-29}$   & \cite{th+ch99}  \\
\hline
\multicolumn{7}{c}{Neutron star/white dwarf binaries} \\
\hline
J0621+1002  & 8.3186813 &       0.00245744      &       0.0116(8) &
$9.288 {\times} 10^{24} \Lambda /(1~\mathrm{km}^{-2})$       & $9 {\times} 10^{-29}$ & \cite{spl+al02}
\\
J1141-6545 & 0.171876 & 0.1976509587    &       5.3084(9)  & $1.881
{\times} 10^{23}\Lambda /(1~\mathrm{km}^{-2}) $       & $5 {\times} 10^{-27}$ & \cite{wil05}
\\
J1713+0747  & 67.82512987 & 0.0000749406    &
0.0006(4)\footnotemark[1]
  &       $7.573{\times} 10^{25} \Lambda /(1~\mathrm{km}^{-2}) $        &       $8 {\times} 10^{-30}$ &
\cite{spl+al05}
\\
B1802-07    &       2.617           &       0.212           & 0.0578(16) &
$2.856{\times} 10^{24}\Lambda /(1~\mathrm{km}^{-2}) $ & $ 6 {\times} 10^{-28}$ &
\cite{th+ch99}
\\
J1906+0746 & 0.085303(2) & 0.165993045(8) & 7.57(3) &
  $9.392{\times}10^{22}  \Lambda /(1~\mathrm{km}^{-2})$ & $ 3 {\times} 10^{-25}$ & \cite{lor+al05}
\\
\hline
\multicolumn{7}{c}{Double pulsars} \\
\hline
J0737-3039  & 0.087779 & 0.102251563     &       16.90(1) &
$9.750{\times} 10^{22} \Lambda /(1~\mathrm{km}^{-2}) $  & $1 {\times} 10^{-25}$ & \cite{wil05}
\\
\hline
\multicolumn{7}{c}{Unknown companion} 
\\
\hline
B1820-11  & 357.7622(3)    &      0.79462(1) & 0.01\footnotemark[1] &
$2.425 {\times} 10^{26}  \Lambda /(1~\mathrm{km}^{-2}) $ &  $4 {\times} 10^{-29}$ & \cite{ly+mc89}
\\
\end{tabular}
\end{ruledtabular}
\footnotetext[1]{Upper limit}
\end{table}

Binary pulsars have been providing unique possibilities of probing
gravitational theories. Relativistic corrections to the binary
equations of motion can be parameterized in terms of post-Keplerian
parameters \cite{wil93}. As seen before, the advance of periastron of
the orbit, $\dot{\omega}$, depends on the total mass of the system and
on the cosmological constant. In principle, because Keplerian orbital parameters
such as the eccentricity $e$ and the orbital period $P_\mathrm{b}$ can
be separately measured, the measurement of $\dot{\omega}$ together
with any two other post-Keplerian parameters would provide three
constraints on the two unknown masses and on the cosmological
constant. As a matter of fact for real systems, the effect of $\Lambda$ is much smaller
than $\dot{\omega}_{\mathrm{pN}}$, so that only upper bounds on the
cosmological constant can be obtained by considering the uncertainty
on the observed periastron shift. We considered binary systems with
measured periastron shift, see Table~\ref{tab:1}. The effect of $\Lambda$
is maximum for B1820-11 and PSR J1713+0747. Despite of the low accuracy in the
measurement of $\dot{\omega}$, PSR J1713+0747 provides the best
constraint on the cosmological constant, $\Lambda \ls 8
{\times}10^{-30}\mathrm{km}^{-2}$. Uncertainties as low as $\delta\dot\omega
\gs 10^{-6}$ have been achieved for very well observed systems, such
as B1913+16 and B1534+12. Such an accuracy for B1820-11 would allow
to push the bound on $\Lambda$ down to $10^{-33}\mathrm{km}^{-2}$.

Better constraints could be obtained by determining post-Keplerian
parameters in very wide binary pulsars. We examined systems with known
period and eccentricity as reported in \cite{lor05}. The binary pulsar
having the most favourable orbital properties for better constraining
$\Lambda$ is the low eccentricity B0820+02, located in the Galactic
disk, with $\dot{\omega}_\Lambda
\sim 1.4{\times}10^{27}\Lambda / (1~\mathrm{km}^{-2})\deg/\mathrm{days}$. For binary pulsars J0407+1607, B1259-63, J1638-4715 and J2016+1948,
the advance of periastron due to the cosmological constant is between
7 and $9{\times}10^{26}\Lambda/ (1~\mathrm{km}^{-2}) \deg/\mathrm{days}$. All of these shifts are
of similar value or better than the Mars one. A determination of
$\dot{\omega}$ for B0820+02 with the accuracy obtained for B1913+16, i.e.
$\delta \dot{\omega}\gs~10^{-6} \deg/\mathrm{days}$  would allow to
push the upper bound down to $ 10^{-34}-10^{-33}\mathrm{km}^{-2}$.

\section{Conclusions}
\label{sec:concl}

We considered the $N$-body equations of motion in presence of a
cosmological constant. The impact of  $\Lambda$ on the two-body system
was explicitly derived. Due to the anti-gravity effect of the
cosmological constant, the barycentre of the system drifts away. The
relative motion is like that of a test particle in a Schwarzschild-de
Sitter space-time with a source mass equal to the total mass of the
two-body system. The main effect of $\Lambda$ is the precession of the
pericentre on the orbital motion.

We determined observational limits on the cosmological constant
from measured periastron shifts. With respect to previous
similar analyses performed in the past on solar system planets,
our estimate was based on a recent determination of the planetary
ephemerides properly accounting for the quadrupole moment of the Sun
and for major asteroids. The best constraint comes from Mars and Earth,
$\Lambda \ls 1-2\times 10^{-36}\mathrm{km}^{-2}$.

Due to the experimental accuracy, observational limits on
$\Lambda$ from binary pulsars are still not competitive with
results from interplanetary measurements in the solar system. Accurate
pericentre advance measurements in wide systems with orbital periods
$\gs 600~\mathrm{days}$ could give an upper bound of $\Lambda \ls
10^{-34}-10^{-33}\mathrm{km}^{-2}$, if determined with the accuracy performed for B1913+16, i.e.
$\delta\dot\omega \gs 10^{-6}\deg/\mathrm{years}$. For some binary pulsars, observations with an accuracy comparable to that achieved in the solar system could allow to get an upper limit on $\Lambda$ as precise as one obtains from Mars data.

The bound on $\Lambda$ from Earth or Mars perihelion shift is nearly $\sim
10^{10}$ times weaker than the determination from observational cosmology, $\Lambda
\sim 10^{-46}\mathrm{km}^{-2}$, but it still gets some relevance. The
cosmological constant might be the non perturbative trace of some
quantum gravity aspect in the low energy limit \cite{pad05}. $\Lambda$
is usually related to the vacuum energy density, whose properties
depends on the scale at which it is probed \cite{pad05}. So that, in
our opinion, it is still interesting to probe $\Lambda$ on a scale of
order of astronomical unit. Measurements of periastron shift should be
much better in the next years. New data from space-missions should get
a very high accuracy and might probe spin effects on the orbital
motion \cite{oco04,ior05}. A proper consideration of the
gravito-magnetic effect in these analyses plays a central role to
improve the limit on $\Lambda$ by several orders of magnitude.

{\it Note added}. After submission of this work, L.~Iorio \cite{ior05b} presented an analysis of solar system data similar to our results in section~\ref{sec:obserA}.

\begin{acknowledgments}
The authors thank N. Straumann for stimulating discussions. M.S. is
supported by the Swiss National Science Foundation and by the Tomalla
Foundation.
\end{acknowledgments}


\end{document}